\documentclass[twocolumn,english,prl,showpacs,superscriptaddress]{revtex4}
\usepackage[colorlinks=true,urlcolor=blue,citecolor=blue,linkcolor=blue]{hyperref}
\usepackage[sort&compress]{natbib}
\usepackage[T1]{fontenc}
\usepackage[latin9]{inputenc}
\usepackage{amssymb}
\usepackage{graphicx}
\usepackage{amsmath,color}
\usepackage{mathrsfs}
\usepackage{float}
\usepackage{indentfirst}
\usepackage{babel}
\usepackage{bbold}

\makeatletter

\begin{document}

\title{Thermodynamics and magnetic properties of the anisotropic 3D Hubbard model}

%

\author{Jakub Imri\v ska}
\affiliation{Theoretische Physik, ETH Zurich, 8093 Zurich, Switzerland}

\author{Mauro Iazzi}
\affiliation{Theoretische Physik, ETH Zurich, 8093 Zurich, Switzerland}

\author{Lei Wang}
\affiliation{Theoretische Physik, ETH Zurich, 8093 Zurich, Switzerland}

\author{Emanuel Gull}
\affiliation{University of Michigan, Ann Arbor, MI 48109, USA}

\author{Daniel Greif}
\affiliation{Institute for Quantum Electronics, ETH Zurich, 8093 Zurich, Switzerland}

\author{\\Thomas Uehlinger}
\affiliation{Institute for Quantum Electronics, ETH Zurich, 8093 Zurich, Switzerland}

\author{Gregor Jotzu}
\affiliation{Institute for Quantum Electronics, ETH Zurich, 8093 Zurich, Switzerland}

\author{Leticia Tarruell}
\affiliation{Institute for Quantum Electronics, ETH Zurich, 8093 Zurich, Switzerland}
\affiliation{LP2N UMR 5298, Univ. Bordeaux 1, Institut d'Optique and CNRS, 33405 Talence, France}
\affiliation{ICFO-Institut de Ciencies Fotoniques, Mediterranean Technology Park, 08860 Castelldefels (Barcelona), Spain}
\author{Tilman Esslinger}
\affiliation{Institute for Quantum Electronics, ETH Zurich, 8093 Zurich, Switzerland}
\author{ Matthias Troyer}
\affiliation{Theoretische Physik, ETH Zurich, 8093 Zurich, Switzerland}

\begin{abstract}

We study the 3D Hubbard model with anisotropic nearest neighbor tunneling amplitudes using the dynamical cluster approximation and compare the results with a quantum simulation experiment using ultracold fermions in an optical lattice, focussing on magnetic correlations. We find that the short-range spin correlations are significantly enhanced in the direction with stronger tunneling amplitudes. Our results agree with the experimental observations and show that the experimental temperature is lower than the strong tunneling amplitude. We characterize the system by examining the spin correlations beyond neighboring sites and determine the distribution of density, entropy and spin correlation in the trapped system. We furthermore investigate the dependence of the critical entropy at the N\'eel transition on anisotropy.
\end{abstract}



\maketitle


The Hubbard model is one of the simplest condensed matter models incorporating the complex interplay between the itinerant and localized behavior of fermions on a lattice. Its phase diagram is expected to contain a number of interesting phases, such as pseudo-gap states, magnetic long-range order and $d$-wave superconductivity~\cite{Maier:2000ds,PhysRevLett.86.139,PhysRevLett.95.237001,Gull:2013hhb,Scalapino,Schulz1996}. Capturing the entire phase diagram theoretically has turned out to be a challenging task, where no unbiased numerical method exists in the interesting strongly correlated region. Furthermore, a detailed validation of theoretical results by comparison to measurements in real materials is often hindered by their structural complexity and limited knowledge of their system parameters.

In this context, the controlled setting of ultracold fermions in optical lattices offers the possibility to directly realize the Hubbard model~\cite{Esslinger:2010ex,Lewenstein:2007p32635} in an experiment and has allowed for studying the metal to Mott-insulator crossover~\cite{Jordens:2008im, Schneider:2008is}. While the low-temperature phase diagram of the Hubbard model has so far not been accessed experimentally, short-range quantum magnetism has been observed in a recent experiment~\cite{Greif:2013kb}. In particular, anti-ferromagnetic spin correlations on neighboring sites were measured using an anisotropic simple cubic lattice configuration, in which the tunneling along one direction was enhanced. In contrast to previous measurements, where a perturbative high-temperature expansion was sufficient to describe the system~\cite{PhysRevLett.104.180401, Greif:2011fra}, understanding this new quantum simulation experiment requires a more sophisticated theoretical approach. Open questions included the influence of the anisotropy on the temperature of the system and the entropy distribution in the trap.

Although the thermodynamics, spin correlations and N\'eel transition temperature for the isotropic 3D Hubbard model have been calculated with different numerical methods ~\cite{Staudt:vz,Jarrell:2005ec,Fuchs:2011ch,PhysRevLett.107.086401,Kozik:2013nt}, the anisotropic Hubbard model was only studied in the Heisenberg limit \cite{PhysRevLett.94.217201}. General questions of the anisotropic Hubbard model concern the strength, range and orientation of spin correlations, as well as the role of dimensionality. In this Letter we perform a quantitative analysis of the anisotropic Hubbard model using the dynamical cluster approximation (DCA)~\cite{Maier:2005tj} with careful extrapolations and compare it to results of a quantum simulation experiment using ultracold fermions in an optical lattice, including additional data~\cite{Greif:2013kb}. To take into account the trapping potential in the experiment, we use the local density approximation (LDA), which has been proven to be accurate in the temperature region relevant for comparison with the experiment~\cite{Scarola:2009cb, Zhou:2011dj}. 
The calculated and experimentally measured spin correlations are found to be in good agreement for temperatures down to the tunneling energy, showing a strong enhancement for large tunneling anisotropies.

\begin{figure}[t]
\centering
\includegraphics[width=8cm]{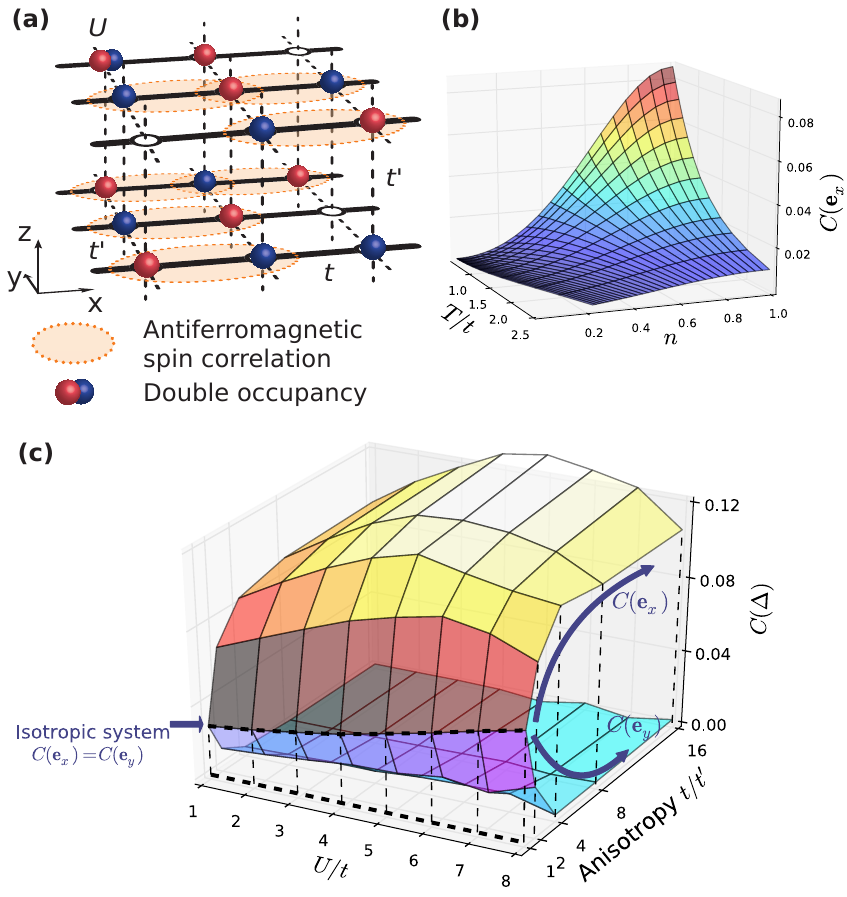}
\caption{(a) Sketch of the anisotropic 3D Hubbard model according to Eq.~(\ref{eq:Ham}). (b)  N.n. spin correlation $C({\bf e}_x)$ vs. filling and temperature for $t/t^{\prime}=7.36$, $U=1.4375t$ in a homogeneous system. (c) N.n. spin correlation along the strong tunneling $C({\bf e}_x)$ (upper surface) and in the transverse direction $C({\bf e}_y)$ (lower surface) for a homogeneous system at half-filling and $T=0.5t$ as a function of anisotropy and interaction strength.}
\label{fig:EOS}
\end{figure}

The Hamiltonian of the anisotropic Hubbard model on a cubic lattice is given by
\begin{eqnarray}
	\hat{H}&=& -t \sum_{{\bf r},\sigma} \left( \hat{c}^\dag_{{\bf r}+{\bf e}_x\sigma} \hat{c}_{{\bf r}\sigma} + h.c.\right) 
	 \nonumber\\
	&&  -t^\prime \sum_{{\bf r},\sigma} \left( \hat{c}^\dag_{{\bf r}+{\bf e}_y\sigma} \hat{c}_{{\bf r}\sigma} + \hat{c}^\dag_{{\bf r}+{\bf e}_z\sigma} \hat{c}_{{\bf r}\sigma} + h.c. \right) \nonumber \\ 
 && +U \sum_{\bf r} \hat{n}_{{\bf r}\uparrow} \hat{n}_{{\bf r}\downarrow}  -\mu \sum_{{\bf r},\sigma} \hat{n}_{{\bf r}\sigma} ,
	 \label {eq:Ham}
\end{eqnarray}
where $\hat{c}^\dag_{{\bf r}\sigma}$ ($\hat{c}_{{\bf r}\sigma}$) creates (annihilates) a fermion at lattice site ${\bf r}$ with spin $\sigma\in\left\{\uparrow,\downarrow\right\}$; $\hat{n}_{{\bf r}\sigma}\equiv \hat{c}^\dag_{{\bf r}\sigma} \hat{c}_{{\bf r}\sigma}$ denotes the occupation number operator; ${\bf e}_i$ denotes the unit vector (setting the lattice spacing to $1$) along the direction $i\in\left\{x,y,z\right\}$. The system has the tunneling amplitude $t$ along the $x$-axis and $t'$ in the directions $y$, $z$ as shown in Fig.~\ref{fig:EOS}(a). The repulsive on-site interaction energy is denoted by $U>0$ and the chemical potential by $\mu$. The ratio $t/t'$ will be referred to as the anisotropy of the system. In this paper, we consider $t/t'\geq 1$, covering the range from an isotropic 3D system to weakly coupled 1D chains.


We study the physical properties of Eq.(\ref{eq:Ham}) with DCA,  
using the numerically exact continuous time auxiliary field quantum Monte Carlo impurity solver~\cite{Gull:2008cm,Gull:2011hh}. The clusters were chosen bipartite and prolongated in the strong tunneling direction proportionally to the anisotropy. The DCA method with extrapolation in cluster size supplies exact results in the thermodynamic limit~\cite{Jarrell:2002}.

We have calculated the thermodynamic properties including energy ($e$) and density ($n$) per site at a given chemical potential $\mu$ and the inverse temperature $\beta=1/T$ (setting $k_{B}=1$). Tabulated equation of state (EOS)~\footnote{In regions with very low filling we have obtained the EOS employing the Hartree approximation.} data may be found in the supplemental materials~\cite{supplementary}. Owing to particle-hole symmetry, the EOS for $n>1$ can be easily determined using the data for $2-n$. The entropy per site $s(\beta)$ is obtained by numerical integration 
\begin{eqnarray}
	s(\beta) &=& s(\beta_0) + f(\beta) \beta - f(\beta_{0}) \beta_0 - \int_{\beta_0}^\beta f(\beta') \:\mathrm{d}\beta' ,\label{eq:entropy_integration}
\end{eqnarray}
with $f(\beta)=e(\beta)-\mu n(\beta)$. The value of $s(\beta_{0})$ at a sufficiently small $\beta_0\sim\frac{1}{50t}$ was obtained using the high temperature series expansion (HTSE) 
\begin{eqnarray}
	s(\beta_{0}) &=& \ln 4 - \frac{\beta_{0}^2}{2}\left[\frac{U^2}{16}+\frac{(\mu-U/2)^2}{2}+t^2+2t^{\prime 2} \right]  \nonumber\\
	&& + \frac{\beta_{0}^3}{8} U \left(\mu-\frac{U}{2}\right)^2 + O(\beta_{0}^4) .
\end{eqnarray}

In addition to the thermodynamic properties, we calculate the equal-time spin correlation function
\begin{eqnarray}
	C({\bf \Delta}) &=& -2\sum_{\bf r}\left\langle \hat{S}^z_{\bf r} \hat{S}^z_{{\bf r}+{\bf \Delta}}\right\rangle ,
\end{eqnarray}
where $\hat{S}^z_{\bf r} = \frac{1}{2}(\hat{n}_{{\bf r}\uparrow} -  \hat{n}_{{\bf r}\downarrow})$ and ${\bf \Delta}$ is a lattice vector. Fig.~\ref{fig:EOS}(b) shows $C({\bf e}_x)$ for various fillings and temperatures at fixed $t/t'=7.36$ and $U=1.4375t$, which was used in the experiment of Ref.~\cite{Greif:2013kb}. Antiferromagnetic correlations between nearest neighbors (n.n.) correspond to positive values of $C({\bf e}_x)$. The signal is greatly enhanced for $T \lessapprox t$ and close to half-filling. At fixed temperature and interaction strength, the n.n. spin correlation along the longitudinal direction $C({\bf e}_x)$ is enhanced with anisotropy $t/t'$, while the correlation along the transverse direction $C({\bf e}_y)$ is suppressed, see Fig.~\ref{fig:EOS}(c). 
$T/t'$ is higher in the anisotropic case and thus the development of spin correlations in the transverse direction $y$ is suppressed. At the same time $C({\bf e}_x)$ is enhanced because singlet formation is facilitated by the effective lowering of dimensionality \cite{PhysRevA.85.061602}. This in turn is caused by the difference in the relevant energy scales: $T$ and $t$ are of the same order but an order of magnitude larger than $t^\prime$~

The quantum simulation experiment is performed as described in detail in Ref.~\cite{Greif:2013kb} using a balanced spin-mixture of the $m_{F}=-9/2,-7/2$ sublevels of the $F=9/2$ hyperfine manifold of $^{40}$K. About $60,000$ fermions are prepared at 10\% of the Fermi temperature in a harmonic optical dipole trap. The gas is then heated to control the entropy per particle in the  trap, which is measured using fits to a Fermi-Dirac distribution. After setting the $s$-wave scattering length to $106(1)$ Bohr radii, an anisotropic cubic optical lattice operating at a wavelength of $\lambda=1064\:\mathrm{nm}$ is turned on using an S-shaped ramp lasting $200\:\mathrm{ms}$. The parameters of the Hubbard model describing the final lattice configuration are computed using Wannier functions.

In order to detect the number of singlets and triplets consisting of two neighboring atoms with opposite spins, we suddenly ramp to a deep simple cubic lattice, suppressing all tunnelings. 
A magnetic field gradient is then used to induce coherent oscillations between singlet and triplet states. 
Subsequently, we merge neighboring sites adiabatically using a tunable-geometry optical lattice~\cite{Tarruell:2012db} and detect the number of double occupancies in the lowest band created by merging. The difference between the fraction of atoms detected in a singlet ($p_{\mathrm{s}}$) or triplet ($p_{\mathrm{t_0}}$) configuration can be used to compute the spin correlator
\begin{equation}
-\langle S^{x}_{\textbf{r}}S^{x}_{\textbf{r}+\textbf{e}_x}\rangle-\langle S^{y}_{\textbf{r}}S^{y}_{\textbf{r}+\textbf{e}_x}\rangle=\left(p_{\mathrm{s}}-p_{\mathrm{t_0}}\right)/2,
\label{eq:PsPtSxSy}
\end{equation}
which is equal to $C({\bf e}_x)$ given the SU(2) invariance of the Hubbard model Eq.(\ref{eq:Ham}).

\paragraph{Results and Discussions} In order to compare the extrapolated DCA data results (obtained for a homogeneous system) with the experiment (performed in a harmonic trap), we use LDA. In LDA, the local quantities at each position ${\bf r}$ in a system with position-dependent chemical potential $\mu({\bf r})$ are approximated by the corresponding quantities in a homogeneous system at $\mu\equiv\mu({\bf r})$ and using the same (trial) temperature. The chemical potential in the calculation is quadratic, $\mu(r)= \mu_{0} -\frac{1}{2}m \bar{\omega}^{2}r^{2}$, where $\bar{\omega}$ is the geometric mean of the trapping frequencies taken from the experiment, $r$ the normalized distance from the trap center and $m$ the atomic mass of $^{40}\rm{K}$. By tuning of the chemical potential in the center $\mu_{0}$ we can match the experimentally measured total particle number $N$ in the trap. We then calculate the entropy and n.n. spin correlations averaged over the trap to compare with the experimentally measured quantities. 

\begin{figure}[t]
\centering
\includegraphics[width=7.5cm]{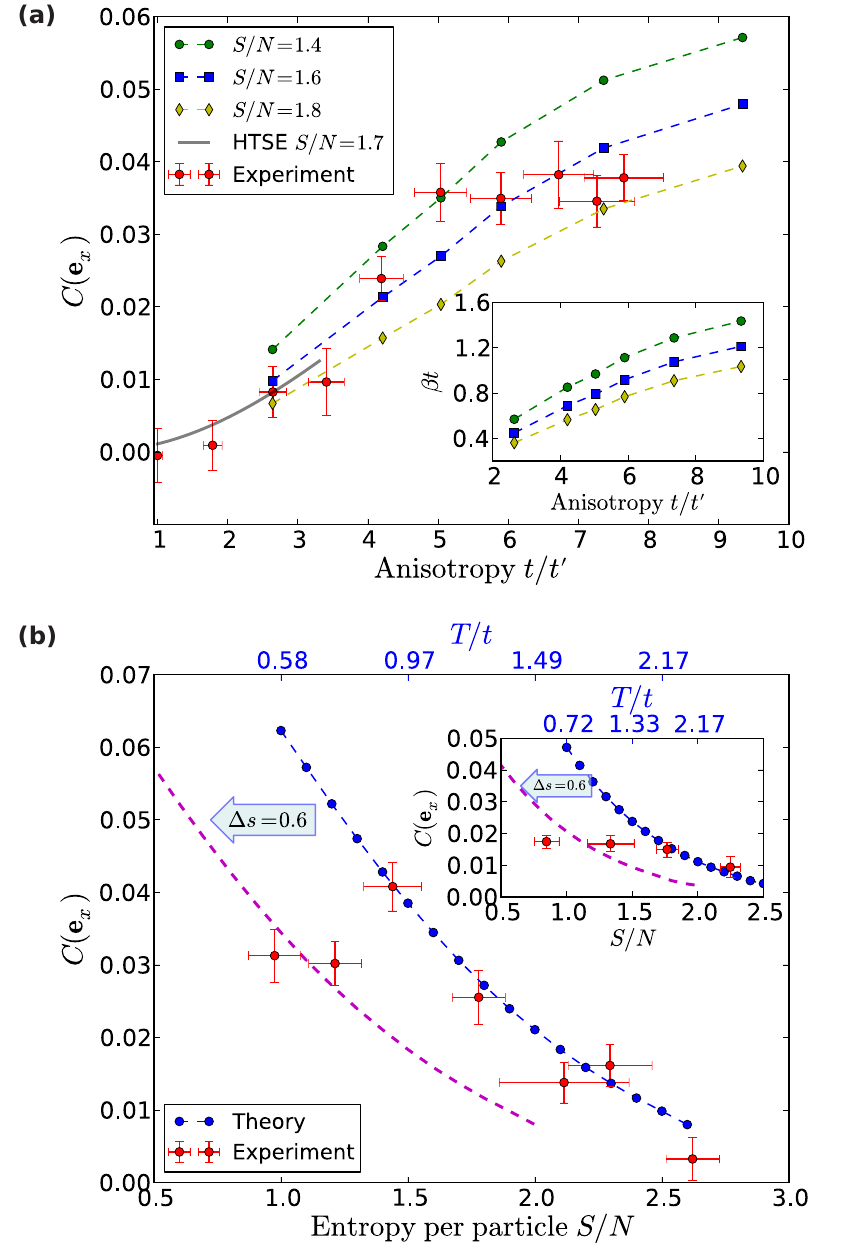}
\caption{Comparison of the calculated spin correlations from DCA+LDA with the experiment. (a)~N.n. spin correlation for different anisotropies and interaction strengths. The entropy per particle before loading into the lattice is below $1.0$ in the experiment; further detailed parameters are listed in the supplemental materials~\cite{supplementary}. Theoretical calculations with different entropies per particle are shown as symbols connected by dashed lines. The solid line shows HTSE results with $S/N=1.7$. The inset shows the inverse temperature $\beta t$ versus anisotropy used in the DCA+LDA calculations. (b)~N.n. spin correlation as a function of entropy per particle for $t/t^{\prime} = 7.36$ and $U = 1.4375t$. The experimental data is plotted as a function of the initial $S/N$ before loading into the lattice, and the blue curve is the theoretical prediction. The upper axis denotes the corresponding temperature determined from the DCA+LDA calculation. For the lowest initial entropies the measured spin correlation deviates from the expected value. This could be attributed to heating during lattice loading. These experimental data points agree with the purple curve, corresponding to an entropy increase of $0.6$. The inset shows results for a different set of parameters ($t/t^{\prime}= 4.21$, $U = 2.98t$) and a comparison with previously unpublished experimental data.}
\label{fig:CvsS}
\end{figure}

Fig.~\ref{fig:CvsS}(a) shows the calculated n.n. spin correlation versus anisotropy together with the experimental data. We find good agreement between the DCA+LDA calculation and the experimental data assuming an entropy per particle $S/N$ in the range of $1.4$ to $1.8$. For anisotropies $\gtrapprox 5$ the experiment enters a regime where corrections to the single band Hubbard model Eq.(\ref{eq:Ham}) may start to play a role in the shallow optical lattice~\cite{Georges}. Close to the isotropic limit, the second order HTSE with $S/N=1.7$ describes the data well. For increasing anisotropies, the HTSE becomes unreliable as the expansion parameter $\beta t$ reaches one. The inset of Fig.~\ref{fig:CvsS}(a) shows that the introduction of the anisotropy leads to a situation where the temperature becomes comparable to or lower than the strong tunnel coupling $t$. The average $C({\bf e}_x)$ increases monotonously with anisotropy, which is a consequence of both the enhancement of correlations for a given $\beta t$ and additionally the increasing $\beta t$.

For a fixed anisotropy $t/t'=7.36$, Fig.~\ref{fig:CvsS}(b) shows the trap averaged $C({\bf e}_x)$ versus entropy per particle (for the experimental data the horizontal axis denotes the initial entropy per particle measured before loading into the lattice). Without any free parameters and assuming no heating, we find excellent agreement for entropies of $1.4k_{\mathrm{B}}$ and above, showing that magnetic effects in the Hubbard model can be accurately studied in this regime. For lower entropies, the experimentally measured spin correlation does not increase further, deviating from the theoretical prediction. This suggests that additional heating may have occurred during the optical lattice loading process, or the system may not have fully equilibrated in the lattice for the lowest initial entropies. This is an important outcome of this study not deducible from the experimental data alone.
A similar situation is found in previous studies of dimerized and simple cubic optical lattices~\cite{Greif:2013kb, Greif:2011fra}. The inset of Fig.~\ref{fig:CvsS}(b) shows a comparison at a different anisotropy $t/t'=4.21$, where similar agreement at high entropies and deviations at low entropies are found. 

The upper horizontal axis of Fig.~\ref{fig:CvsS}(b) shows the temperature used in the DCA+LDA calculations. 
For the lowest entropy $S/N=1.4$, where the experimentally measured spin correlator matches the theoretical value, the temperature is found to be $T\approx 0.88t$. 
This indicates that an anisotropic 3D lattice is a viable system for an experimental study of the low-temperature regime of the Hubbard model in effectively one dimension~\cite{Giamarchi:2003} at currently accessible experimental entropies. 

\begin{figure}[t]
\centering
\includegraphics[width=8cm]{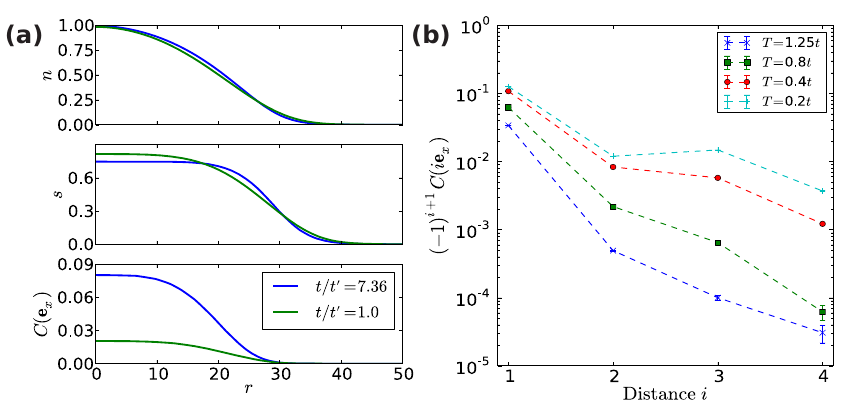}
\caption{(a)~The distribution of density, entropy and n.n. spin-correlation per site in the harmonic trap vs. distance from the center. The simulation is done with $U=1.4375t, N=50000$, the trap averaged entropy is $S/N=1.6$. The trapping frequency is chosen such that the filling is $n=1$ at the trap center. (b)~Extrapolated spin correlations as a function of distance along the $x$ axis in the paramagnetic phase for $t/t'=7.36$, $U=1.44t$ and half-filling for different temperatures. 
\label{fig:characterization}}
\end{figure}

Fig.~\ref{fig:characterization}(a) shows the calculated distribution of the density, entropy and n.n. spin correlation in the trap for the isotropic ($t/t'=1$) and anisotropic ($t/t'=7.36$) Hubbard model with the same $U/t$, particle number and entropy per particle. In each case we set $\mu_{0}=U/2$ to obtain the filling $n=1$ in the trap center and tune $\bar{\omega}$ such that atom number is constant. The corresponding temperatures are $T=0.95t$ and $T=0.58t$ resp~\footnote{$T/t$ is lower in the anisotropic case because of reduction of the total bandwidth}. Owing to qualitatively similar equations of state between the isotropic and anisotropic case at fixed tunneling $t$, we find a very similar behavior for both the density and entropy distribution in the trap. 
This is in contrast with the dimerized lattice examined in~\cite{Greif:2013kb}, which has an energy gap. In Fig.~\ref{fig:characterization}(a) the n.n. spin correlations are more pronounced for large anisotropy when comparing to the isotropic case, similar to the results in Fig.~\ref{fig:EOS}(c). 
To further characterize the state realized in the experiment, we compute the spin correlation beyond n.n. along the $x$ direction, shown in~Fig.~\ref{fig:characterization}(b). It shows an alternating sign with distance, confirming the presence of antiferromagnetic spin correlations ~\footnote{In Fig.~\ref{fig:characterization}(b) we find $\left|{C(3{\bf e}_x)}\right| > \left|{C(2{\bf e}_x)}\right|$ for $T=0.2t$, which is a feature inherited from the half-filled non-interacting system on the cubic lattice, where spin correlations at even Manhattan distances vanish.}. For temperatures close to the lowest experimental temperature the next nearest neighbor (n.n.n.) correlation drops to a value below the current experimental resolution. Its behavior with distance suggests that the correlation decays exponentially.


\begin{figure}[t]
\centering
\includegraphics[width=6cm]{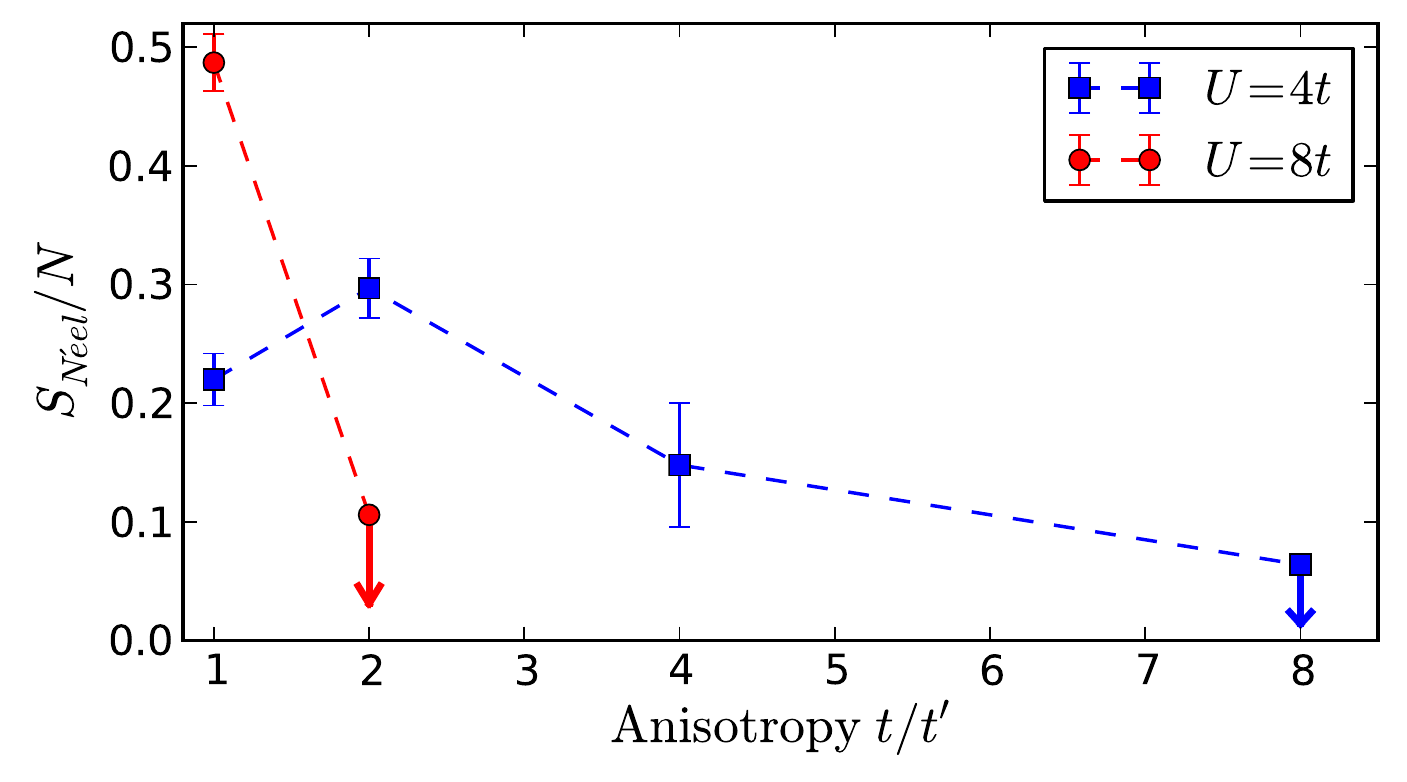}
\caption{Critical entropy per particle $S(T_{\textrm{N\'eel}})/N$ at the N\'eel transition vs. anisotropy for two different interactions at half-filling. The data points shown with an arrow are upper bounds owing to the difficulty of obtaining the  extrapolated $T_{\textrm{N\'eel}}$ or a reliable $s(T)$ down to the extrapolated transition temperature.}
\label{fig:S_Neel}
\end{figure}
Finally, we address the question of how the introduction of anisotropy affects the N\'eel transition in a 3D lattice. We focus on the half-filled system. Fig.~\ref{fig:S_Neel} shows the calculated critical entropy at the N\'eel transition for different anisotropies. The critical entropy at $U=4t$ shows a nonmonotonic behavior as a function of anisotropy. We explain this by the reduction of the total bandwidth $W=4(t+2t')$ and thus by the effective increase of the interaction strength ($U/W$) towards the optimal value $U/W\approx 2/3$ for the isotropic system~\cite{Fuchs:2011ch}. Consistent with our simple argument, the curve for $U=8t$ decays monotonically. We find that the introduction of anisotropy does not enhance the critical entropy over the optimum value for the isotropic case ($S/N\approx 0.487(24)$ in the present study).

The estimate of the N\'eel temperature was obtained for a set of clusters within the DCA simulation by looking for the divergence of the static antiferromagnetic spin susceptibility~\cite{Maier:2005tj}. The $T_{\textrm{N\'eel}}$ was then obtained by extrapolation as suggested in~\cite{Jarrell:2005ec}~\footnote{For $t=t'$ our model is part of the universality class of the 3D $S = 1/2$ Heisenberg model and for $t\neq t'$ it belongs to the classical 3D Heisenberg universality class, both of which have a critical exponent of $\nu\approx 0.71$~\cite{Sandvik1998}.}. Fig.~\ref{fig:S_Neel} shows $s(T_{\textrm{N\'eel}})$ with curve  
$s(T)$ integrated within the paramagnetic phase. Our results for the isotropic case, $T_{\textrm{N\'eel}}=0.194(4)$ for $U=4t$ and $T_{\textrm{N\'eel}}=0.360(9)$ for $U=8t$, are consistent with previous studies~\cite{Staudt:vz,Jarrell:2005ec}. Both estimates are slightly above the estimates $T_{\textrm{N\'eel}}<0.17t$, $T_{\textrm{N\'eel}}=0.3325(65)$ obtained by diagrammatic determinantal Monte Carlo calculations on larger lattices for $U=4t$ and $U=8t$, resp.~\cite{Kozik:2013nt}.

\paragraph{Conclusions} We have computed the properties of the 3D anisotropic Hubbard model in the regime accessed by the quantum simulation experiment. Short-range spin correlations were shown to be enhanced by anisotropy, even when the critical entropy at the N\'eel temperature is reduced. Our theoretical results show good agreement with our experiments, allowing us to characterize this system in detail. In particular, using the nearest-neighbor spin correlation as a thermometer, the experimentally realized temperature was found to reach values below the strong tunneling amplitude. 
Given the access to effectively one-dimensional Hubbard-chains featuring spin order, the tunability of an optical lattice system may be used to probe their excitation dynamics or the crossover from 1D to higher dimensions~\cite{Giamarchi:2003}.


\paragraph{Note} During the preparation of this manuscript, we became aware of a related finite temperature study of the 1D Hubbard model~\cite{Corinna}. 

We thank Jan Gukelberger, Michael Messer and James LeBlanc for useful discussions. This work was supported by the ERC Advanced Grants SIMCOFE and SQMS, the Swiss National Competence Center in Research QSIT and the Swiss National Science Foundation. The calculations used a code based on the ALPS libraries~\cite{ALPS1.3:2007,ALPS2.0:2011} and were performed on the Brutus cluster at ETH Zurich. 

\bibliographystyle{apsrev4-1}
\bibliography{anisohubbard}


\end{document}